\documentclass[12pt,nofootinbib,prd]{revtex4}

\usepackage{amsmath}
\usepackage{graphicx}

\def\mt{\widetilde{m}}
\def\Mt{\widetilde{M}}
\def\mq{\mt_q}
\def\mg{\mt_g}
\def\mi{\mt_i}
\def\mj{\mt_j}
\def\ma{\mt_1}
\def\mb{\mt_2}

\begin{document}

\title{The mass insertion approximation without squark degeneracy}

\author{Guy \surname{Raz}}
\affiliation{Particle Physics Department \\ Weizmann Institute of
  science \\ Rehovot 76100, Israel}
\email{guy.raz@weizmann.ac.il}

\begin{abstract}
  We study the applicability of the mass insertion approximation (MIA)
  for calculations of neutral meson mixing when squark masses are not
  degenerate and, in particular, in models of alignment. We show that
  the MIA can give results that are much better than an order of
  magnitude estimate as long as the masses are not strongly
  hierarchical. We argue that, in an effective two-squark framework,
  $\mq=(\ma+\ma)/2$ is the best choice for the MIA expansion point,
  rather than, for example, $\mq^2=(\ma^2+\mb^2)/2$.
\end{abstract}

\maketitle


\section{Introduction}
\label{sec:introduction}

The mass insertion approximation (MIA) is often used to simplify
expressions involving supersymmetric contributions to flavour
changing neutral current processes from loop
diagrams~\cite{Gabbiani:1989rb,Hagelin:1994tc,Gabbiani:1996hi}. The
simplification is achieved by the replacement of a sum over all
possible internal propagators and the appropriate mixing at the
vertices, with a single (small) off-diagonal mass insertion in a basis
where all gauge couplings are diagonal. The resulting expressions are
formulated in terms of parameters which can be estimated in
various supersymmetric models.

It may seem, naively, that the smallness of the off-diagonal
mass-squared matrix element would justify the approximation. The true
picture, however, is that these off-diagonal elements are the product
of mixing angles at the vertices and mass-squared differences between
intermediate squarks. The MIA, on the other hand, is a Taylor
expansion only with respect to the latter, namely, the mass-squared
difference (we give an exact formulation of these statements in
section~\ref{sec:form-mass-insert}). A small off-diagonal element does
not necessarily imply a small mass difference. Instead, it may be
related to small mixing angles. But then the validity of the MIA is
questionable.

This is exactly the situation in the framework of quark-squark
alignment (QSA) models~\cite{Nir:1993mx,Leurer:1994gy}. In this class
of supersymmetric models the squark masses-squared are all of the same
order of magnitude, a free parameter denoted by $\mt^2$, and the mass
squared differences between them are also of the same order of
magnitude, that is: $|\mi^2-\mj^2|/\mt^2=\mathcal{O}(1)$. Apriori,
this is a problematic situation for using the MIA. Yet, it is
frequently used in the literature.\footnote{In fact, since QSA
  models allow estimates of $|\mi^2-\mj^2|/\mt^2$ but not of the
  individual masses, the MIA provides the best way to derive
  meaningful results.}

We therefore study the validity of the MIA in the context of QSA
models. We confirm that the approximation is applicable and useful for
such models. We also clarify the connection between the (unknown)
details of the squark mass spectrum and the MIA parameters.

The organization of this work is as follows: We formulate the details
of the MIA in section~\ref{sec:form-mass-insert}. The analysis and our
results for non-degenerate squarks masses are presented in
section~\ref{sec:analys-non-degen}.

\section{Formulation of the mass insertion approximation.}
\label{sec:form-mass-insert}

\begin{figure}[tbp]
  \centering
  \includegraphics[width=0.6\columnwidth]{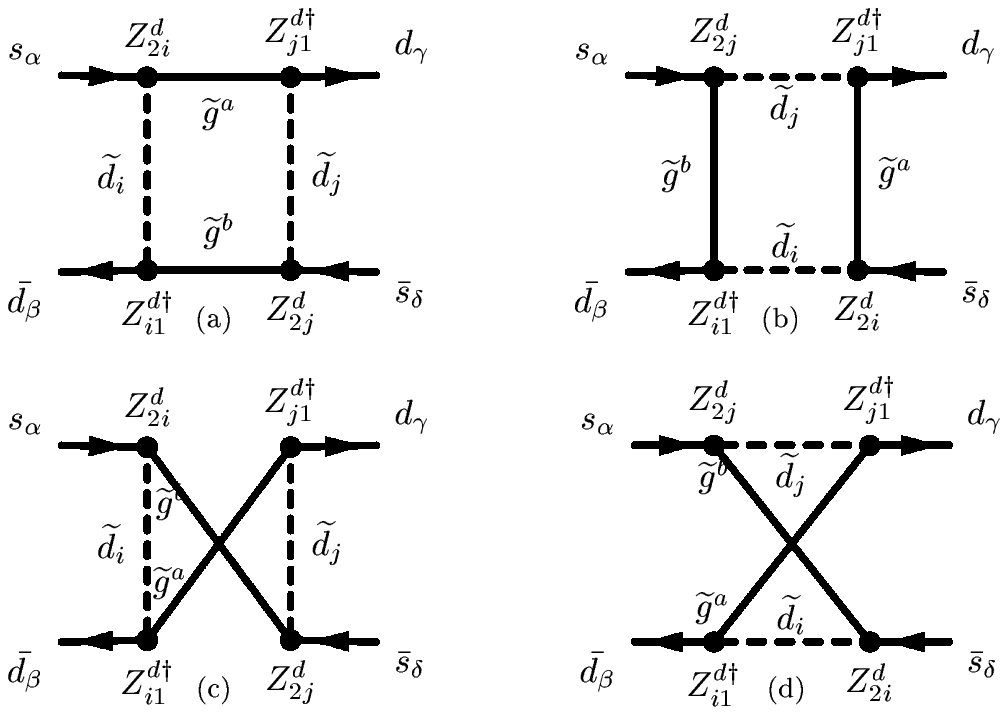}
  \caption{Gluino box diagrams contributing to $K^0$ - $\overline{K^0}$ mixing.}
  \label{fig:boxdiag}
\end{figure}

Let us first formulate the details of the MIA. To illustrate it in the
context of QSA, we study a specific example: The supersymmetric
contribution to neutral $K$ meson mixing from gluino box diagrams with
two intermediate squarks. The relevant diagrams are shown in
figure~\ref{fig:boxdiag}. We focus on the following term arising from
these diagrams~\cite{Nelson:1997bt}:
\begin{equation}
  \label{eq:1}
  M^K_{12} \supset C \left(Z^d_{2i}Z^d{}^\dag_{i1}
    Z^d_{2j}Z^d{}^\dag_{j1} \right) J_4(\mg^2,\mi^2,\mj^2)\;.
\end{equation}
Here $C$ is a numerical factor, given in terms of the $K$ meson parameters:
\begin{equation}
  \label{eq:2}
  C\equiv\frac{{\left(4\pi\right)}^2}{i}\frac{\alpha_s^2 m_K
    f^2_K \hat{B}_K \eta}{2} \simeq \frac{{\left(4\pi\right)}^2}{i}\;
    5.4\times10^4\;\;\;\; \text{MeV}^3\;,
\end{equation}
$Z^d_{ij}$ are the quark-squark mixing angles, and $J_4$ is given
by
\begin{equation}
  \label{eq:3}
  J_4(\mg^2,\mi^2,\mj^2)\equiv \frac{11}{54}
  \widetilde{I}_4(\mg^2,\mi^2,\mj^2)+\frac{2}{27}\mg^2
  I_4(\mg^2,\mi^2,\mj^2)\;,
\end{equation}
with $\mg$ the gluino mass, $\mi,\,\mj$ the down squark masses and
\begin{equation}
    \label{eq:4}
    \begin{split}
      I_4(\mg^2,\mi^2,\mj^2)
      & \equiv \int{\frac{d^4p}{{(2\pi)}^4} \frac{1}
        {(p^2-\mg^2)(p^2-\mg^2)(p^2-\mi^2)
          (p^2-\mj^2)}} \\*
      & = \frac{i}{{\left(4\pi\right)}^2}\Biggl[
      \frac{1}{\left(\mi^2-\mg^2\right)
        \left(\mj^2-\mg^2\right)} + \Biggr. \\*
      & \qquad + \frac{\mi^2}{\left(\mi^2-\mj^2\right)
        {\left(\mi^2-\mg^2\right)}^2}\ln
    \left(\frac{\mi^2}{\mg^2}\right) \\*
    & \qquad \Biggl. +
    \frac{\mj^2}{\left(\mj^2-\mi^2\right)
      {\left(\mj^2-\mg^2\right)}^2}\ln
    \left(\frac{\mj^2}{\mg^2} \right)\Biggr] \;,
  \end{split}
\end{equation}
\begin{equation}
    \label{eq:5}
    \begin{split}
      \widetilde{I}_4(\mg^2,\mi^2,\mj^2)
      & \equiv \int{\frac{d^4p}{{(2\pi)}^4} \frac{p^2}
        {(p^2-\mg^2)(p^2-\mg^2)(p^2-\mi^2)
          (p^2-\mj^2)}} \\*
      & = \frac{i}{{\left(4\pi\right)}^2}\Biggl[
      \frac{\mt^2_g}{\left(\mi^2-\mg^2\right)
        \left(\mj^2-\mg^2\right)} + \Biggr. \\*
      & \qquad + \frac{\mi^4}{\left(\mi^2-\mj^2\right)
        {\left(\mi^2-\mg^2\right)}^2}\ln
    \left(\frac{\mi^2}{\mg^2}\right) \\*
    & \qquad \Biggl. +
    \frac{\mj^4}{\left(\mj^2-\mi^2\right)
      {\left(\mj^2-\mg^2\right)}^2}\ln
    \left(\frac{\mj^2}{\mg^2} \right)\Biggr]\;.
  \end{split}
\end{equation}

The MIA is nothing more than a Taylor expansion of $J_4$. We choose to
expand the squark masses around some point $\mq$. Owing to the
specific form of $J_4$ we find
\begin{equation}
  \label{eq:6}
   J_4(\mg^2,\mi^2,\mj^2) = \sum_{m,n = 0}^\infty
   \frac{C_{m+n}(x)}{\mq^{2(1+m+n)}}          
  \left(\Delta\mi^2\right)^m\left(\Delta\mj^2\right)^n \;,
\end{equation}
where $x\equiv \mg^2/\mq^2$, and $\Delta\mi^2 \equiv
\mi^2-\mq^2$. The coefficient $C_{m+n}(x)/\mq^{2(1+m+n)}$ is the
$(m+n)'$th derivative of $J_4$ evaluated at $\mq$, times the symmetry factor
$1/(m!n!)$. (The exact form of the coefficient is not important for
our purpose.)

Substituting~\eqref{eq:6} in~\eqref{eq:1} we get
\begin{equation}
  \label{eq:7}
  C \sum_{m,n = 0}^\infty
  \left\{\frac{C_{m+n}(x)}{\mq^{2(1+m+n)}}          
    \left[Z^d_{2i}\left(\Delta\mi^2\right)^mZ^{d\dag}_{i1}\right]
    \left[Z^d_{2j}\left(\Delta\mj^2\right)^n 
  Z^{d\dag}_{j1}\right]\right\} \;.
\end{equation}
Note that a sum over $i$ and $j$ is implied. If the MIA is valid, this
expansion converges fast and we can keep only the lowest order terms.
However, due to the unitarity of $Z^d$, the terms with either $m=0$ or
$n=0$ vanish. The first non-vanishing contribution, therefore, will be
from the term with $m=n=1$. This term, however, is special since we
can write (again, a sum over $j$ is implied):
\begin{equation}
  \label{eq:8}
  Z^d_{2j}\left(\Delta\mj^2\right) Z^{d\dag}_{j1} =
  Z^d_{2j}\left(\mq^2+\Delta\mj^2\right) Z^{d\dag}_{j1} = (\Mt^2_d){}_{21}\;,
\end{equation}
where $\Mt^2_d$ is the squark mass-squared matrix in the basis where
quarks masses and gluino couplings are diagonal. Thus, if the MIA holds we can
replace~\eqref{eq:7} with:
\begin{equation}
  \label{eq:9}
  C \times\frac{C_{2}(x)}{\mq^{6}}          
    \left((\Mt^2_d){}_{21}\right)^2 \;.
\end{equation}
We stress that a small $(\Mt^2_d){}_{21}$ is not enough, by itself,
to justify the use of the MIA. The question of validity should be
considered in the context of eq.~\eqref{eq:6}. We note, however,
that the quality\ of the MIA is not completely equivalent to the
quality of the approximation that is obtained by keeping only the lowest terms
in~\eqref{eq:6}. The reason is that the zeroth order term ($m=n=0$), the
first order terms ($m=1,\,n=0$ and $m=0,\,n=1$) and some of the second
order terms ($m=2,\,n=0$ and $m=0,\,n=2$), while appearing
in~\eqref{eq:6}, do not contribute to the mixing amplitude
in~\eqref{eq:7} due to the unitarity of $Z^d$. In other words, the
validity of~\eqref{eq:9} has to do with one of the second order terms
in~\eqref{eq:6}, rather than with all lowest order terms.

Nonetheless, it is obvious that when the expansion parameter is small,
$\left|\Delta\mi^2/\mq^2\right|\ll 1$,
the approximation is good. It is the condition in QSA models, 
$\left|\Delta\mi^2/\mq^2\right|\sim 1$ which needs a special consideration.

\section{The case of non-degenerate masses}
\label{sec:analys-non-degen}

\begin{figure}[tbp]
  \centering
  \includegraphics[width=0.5\columnwidth]{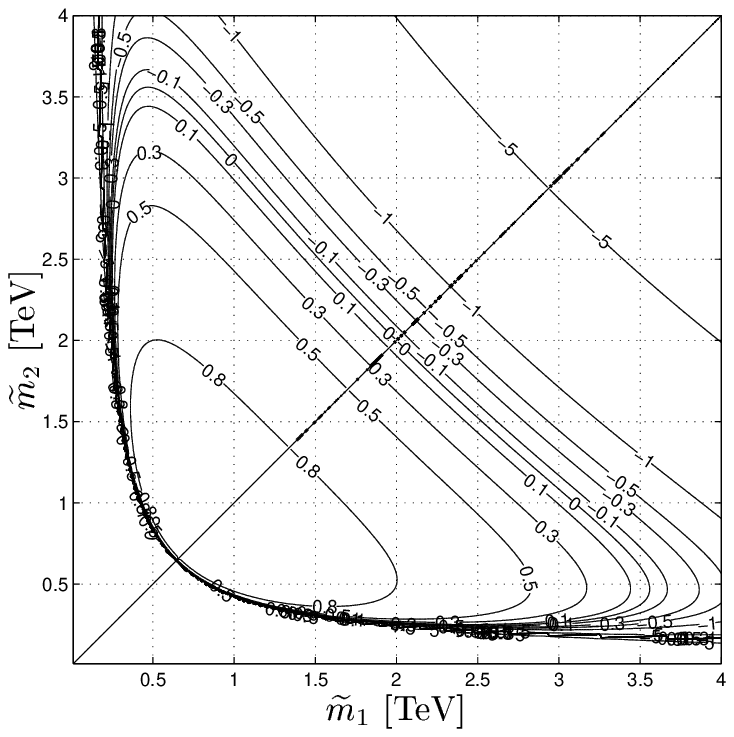}
  \caption{$r$ as a function of $\ma$ and $\mb$ for $\mq=2$ TeV, $\mg=1$ TeV.}
  \label{fig:mqsolve1}
\end{figure}

In order to study quantitatively the non-degenerate case, we
simplify the form of $Z^d$ by assuming that only one mixing angle,
namely $Z^d_{12}$, is large. This is usually the case in QSA
models. Such an assumption allows us to consider
only the first two generations. Expression~\eqref{eq:1} then
simplifies to
\begin{equation}
  \label{eq:10}
  C\times \cos^2 \theta \sin^2 \theta
  \times\left[J_4(\mg^2,\mt_1^2,\mt_1^2)+J_4(\mg^2,\mt_2^2,\mt_2^2)-J_4(\mg^2,\mt_1^2,\mt_2^2)-
    J_4(\mg^2,\mt_2^2,\mt_1^2) \right] \;,
\end{equation}
where $\sin\theta\approx Z^d_{12}$ is the single large mixing
angle. Equivalently,~\eqref{eq:7} can be written as
\begin{equation}
  \label{eq:11}
  C\times \cos^2 \theta \sin^2 \theta \times
  \left(\frac{C_{m+n}(x)}{\mq^{2(1+m+n)}}\right) 
  \times \left( \left(\Delta\ma^2\right)^m -
  \left(\Delta\mb^2\right)^m \right)\left( \left(\Delta\ma^2\right)^n -
  \left(\Delta\mb^2\right)^n \right)\;.
\end{equation}
The above expression manifestly demonstrates the vanishing of terms
with either $m=0$ or $n=0$. The MIA is obtained by keeping
in~\eqref{eq:11} only the term with $m=n=1$
\begin{equation}
  \label{eq:12}
  C\times \cos^2 \theta \sin^2 \theta \times
  \left(\frac{C_2(\mg^2/\mq^2)}{\mq^2}\right) 
  \times \left( \frac{\ma^2 - \mb^2}{\mq^2}\right)^2\;.
\end{equation}
We can now test the accuracy of the MIA by defining the deviation
parameter (using the symmetry of $J_4$ with respect to $\mi^2$ and $\mj^2$)
\begin{equation}
  \label{eq:13}
  r \equiv 1- \frac{\displaystyle \left(\frac{C_2(\mg^2/\mq^2)}{\mq^2}\right) 
    \left( \frac{\ma^2 -
        \mb^2}{\mq^2}\right)^2}{J_4(\mg^2,\mt_1^2,\mt_1^2)+
    J_4(\mg^2,\mt_2^2,\mt_2^2)- 2\, J_4(\mg^2,\mt_1^2,\mt_2^2)}\;. 
\end{equation}
Figure~\ref{fig:mqsolve1} shows $r$ as a function of $\ma$ and $\mb$ for $\mq=2$ TeV, $\mg=1$ TeV.

Obviously, when $\ma\approx\mb\approx\mq$, we expect the deviation to
be small. The interesting result, however, is that the deviation is
small (of order $10\%$) even for non-degenerate $\ma$ and $\mb$, as
long as $(\ma+\mb)/2 \approx \mq$. This is the case, for example, when
$\ma=1$ TeV, $\mb=3$ TeV and $\mq=2$ TeV. In other words, the
result of the exact expression~\eqref{eq:10} using $\ma$ and $\mb$,
can be reproduced using the MIA of~\eqref{eq:12} with
$\mq\approx(\ma+\mb)/2$ and the knowledge of $\Mt_{12}
\sim\cos\theta\sin\theta(\ma^2-\mb^2)/\mq^2$.

We see, therefore, that as long as the squark masses are not strongly
hierarchical, the `right' choice of the expansion point $\mq$ results
with the MIA being a rather accurate approximation. This is very
useful in QSA models where we have $(\ma^2-\mb^2)/\mq^2 =
\mathcal{O}(1)$. The amplitudes in these models can therefore be
expressed in the MIA by using $\mq\approx(\ma+\mb)/2\sim\mt$.
Although this result was demonstrated here for a specific mixing
contribution and using specific assumptions on the mixing angles, we
found it to be quite a general result.\footnote{Moreover,
  the example given here directly applies to the most stringent test
  of QSA models, namely, the contribution to $D^0$ -- $\overline{D^0}$
  mixing.}

The result $\mq\approx(\ma+\mb)/2$ is not trivial. Looking
at~\eqref{eq:11}, it seems that the best strategy is to choose $\mq^2
= (\ma^2+\mb^2)/2$. This choice results with $\Delta\ma^2 =
-\Delta\mb^2$, which eliminates all terms with either $m$ or $n$
even.\footnote{In other words, all terms with $m+n$ odd are
  eliminated, since $m+n$ odd implies either $m$ even and $n$ odd or
  vice versa. Some of the $m+n$ even terms which are due to both $m$
  and $n$ even are eliminated as well.} In particular, it eliminates
the next-to-leading order terms with either $m=1,\,n=2$ or
$m=2,\,n=1$. Naively, one would expect faster convergence in this case
and therefore obtaining a good approximation.

This naive expectation is, however, wrong. Studying the form of the
coefficients $C_{m+n}$ we find that the sign flips between even and
odd $m+n$ terms. The choice $\mq^2 = (\ma^2+\mb^2)/2$, which
eliminates all odd $m+n$ terms, eliminates therefore all the negative
sign terms in~\eqref{eq:11}. Since the $C_{m+n}$ coefficients decrease
slowly, this induces a larger error and a worse approximation.

On the other hand, we find that the choice $\mq\approx(\ma+\mb)/2$ is
the most sensible one since it leads to an approximate cancellation
between the next-to-leading order and the next-to-next-to-leading
order terms. The extent to which such a choice is optimal, as can be
seen in figure~\ref{fig:mqsolve1}, is remarkable.

Although we presented here explicitly only the LL and RR contributions to
the mixing amplitude, the approximation also holds (over a somewhat
smaller range of masses) for the LR and RL contributions.

To summarize, we confirm that the common practice of using the MIA in
supersymmetric models is justified even in the case of non-degenerate
squark masses, as long as the masses are not strongly hierarchical. We
showed that over a wide range of masses, the best strategy is to
choose the MIA expansion point $\mq$ to be the average of the masses
involved. Using this strategy, the MIA provides a surprisingly good
calculation of neutral meson mixing in models where a single mixing
angle dominates, such as QSA~\cite{Nir:1993mx,Leurer:1994gy} and
`effective supersymmetry'~\cite{ Dine:1990jd, Dimopoulos:1995mi,
  Pomarol:1996xc, Cohen:1996vb, Kaplan:1999iq}.

\begin{acknowledgments}
I thank Yossi Nir for his help and comments.  
\end{acknowledgments}



\end{document}